\documentclass{webofc}
\usepackage[varg]{txfonts}   % Web of Conferences font
\usepackage{natbib}
\usepackage[T1]{fontenc} % if needed
\usepackage{comment}
\usepackage{booktabs}
\usepackage{ulem}
\usepackage{subfig}
\usepackage{graphicx}
\usepackage[labelfont=bf,labelsep=period]{caption}
\usepackage{hyperref}

\newcommand{\RTX}{Quadro RTX 8000}
\newcommand{\Tesla}{Tesla V100}
\newcommand{\ifb}{${\rm fb}^{-1}$}
\newcommand{\GeV}{\ensuremath{\rm GeV}}

\begin{document}
\title{Towards a realistic track reconstruction algorithm based on graph neural networks for the HL-LHC}

\author{\firstname{Catherine} \lastname{Biscarat} \inst{1} \and \firstname{Sylvain} \lastname{Caillou} \inst{1} \and \firstname{Charline} \lastname{Rougier} \inst{1} \and \firstname{Jan} \lastname{Stark} \inst{1} \and \firstname{Jad} \lastname{Zahreddine} \inst{1}}

\institute{Laboratoire des 2 Infinis - Toulouse (L2IT-IN2P3), Université de Toulouse, CNRS, UPS,\\ F-31062~Toulouse Cedex 9, France}

\abstract{%
  The physics reach of the HL-LHC will be limited by how efficiently the experiments can use the available computing resources, i.e.
  affordable software and computing are essential.
  The development of novel methods for charged particle reconstruction at the HL-LHC incorporating machine learning techniques or based
  entirely on machine learning is a vibrant area of research. In the past two years, algorithms for track pattern recognition based on
  graph neural networks (GNNs) have emerged as a particularly promising approach. Previous work mainly aimed at establishing proof of principle.
  In the present document we describe new algorithms that can handle complex realistic detectors. The new algorithms are implemented in ACTS, a common framework for tracking software.
  This work aims at implementing a realistic GNN-based algorithm that can be deployed in an HL-LHC experiment. 
}

\maketitle

\flushbottom

%%%%%%%%%%%%%%%%%%%%%%%%%%%%%%%%%%%%%%%%%%%%%
%%%%%%%%%%%%%%%%%%%%%%%%%%%%%%%%%%%%%%%%%%%%%
\section{Introduction}
\label{sec:intro}

The LHC collider and the associated experiments are undergoing a major upgrade that will increase the
size of the datasets of each of the general-purpose experiments ATLAS and CMS by one order of magnitude~\cite{Apollinari:2284929} compared
to the initial LHC plan. These large datasets will enable precise measurements in the Higgs sector.
The original LHC plan is to accumulate 300~\ifb\ of data per experiment. This original
plan is expected to be concluded in 2024, and the installation of the high-luminosity LHC (HL-LHC~\cite{Apollinari:2284929}) is expected to
be completed in 2027~\cite{HLLHCtiming}.
During the HL-LHC phase, each experiment will accumulate at least 3000~\ifb\ of data. The instantaneous luminosity
during the HL-LHC phase will be about three times larger than the highest instantaneous luminosities reached
during the last years of the LHC phase, i.e. the HL-LHC dataset will not only be larger; the events will also be
more complex. The average number of pile-up events is expected to be 200.

The general-purpose detectors ATLAS and CMS will undergo significant upgrades in order to cope with the
complexity of events at the HL-LHC, including the installation of new tracking
subdetectors with improved granularity~\cite{CERN-LHCC-2017-021,CERN-LHCC-2017-005,CERN-LHCC-2017-009}. 
Even with the upgraded detectors, the HL-LHC leads to a steep increase in computing resources needed to process
and analyse the data~\cite{ATLASCDR,CMSnoteHLLHC}. Assuming a flat budget, the expected improvements in computing
hardware will not be able to provide this increase. The amount of data that experiments can collect and process will
be limited by affordable software and computing, and therefore the physics reach during HL-LHC will be limited by how
efficiently these resources can be used~\cite{HSFWhitePaper}.

The offline event reconstruction represents a large fraction of computing resource
needs~\cite{ATLASCDR,CMSnoteHLLHC}. The CPU needed for event reconstruction tends to be dominated by charged particle
reconstruction (tracking)~\cite{HSFWhitePaper,ATLASCDR,CMSnoteHLLHC}. Significant effort is therefore being invested
into the reduction of computing resources needed for tracking. This includes improvements of existing algorithms and the
development of new algorithms, possibly incorporating machine learning~(ML) for track pattern recognition~\cite{HSFWhitePaper}.
To boost the development of ML techniques for tracking and strengthen the involvement of ML~experts in this effort, a
challenge on the Kaggle platform has been organised in 2018-2019: the TrackML Challenge~\cite{Kiehn:2019tbl}.
At the end of the challenge, ML was not at the core of the fastest algorithms~\cite{Rousseau:2020rnz}, and in general
the proposed solutions included less innovative ML than expected. % cite Kiehn at Pascal ?
The authors of Ref.~\cite{Farrell:2018cjr} identify the use of graph neural networks (GNNs)~\cite{battaglia2018relational} as a promising (their most promising)
ML~solution for charged particle tracking at the HL-LHC. Since then, significant effort has been invested in the
development of algorithms based on GNNs~\cite{exatrkxpaper,Choma:2020cry}. 
Promising performance of GNN-based algorithms on the TrackML dataset has been shown in Ref.~\cite{exatrkxpaper}.
The algorithms and code in Ref.~\cite{exatrkxpaper} aim at establishing proof of principle. No attempt at an
implementation for a real-world experiment is made: only the very central part of the detector is considered, this
limited region of the detector is split into 16 sections to avoid any issues related to GPU memory consumption
during the training of the GNN, the algorithm used to construct the graph representation of the experimental data
only works for trivial detector geometries, and the implementation does not aim at making efficient use of memory
or compute power.

In the present document we report initial results from a new effort to implement a realistic GNN-based algorithm that
can be deployed in an HL-LHC experiment. We present a novel algorithm for graph construction that can handle the complex
geometry of a realistic detector, including full coverage in pseudo-rapidity~($\eta$), methods for memory management that allow
GNN training on the full detector without any sectioning, and initial studies of GNNs trained on the full detector.
The algorithms are implemented in the ACTS framework~\cite{Ai:2019kze} to facilitate their use with detailed simulation studies of
HL-LHC detectors, and their direct comparison to other tracking algorithms, e.g. tracking solutions based on a Kalman
filter.

%%%%%%%%%%%%%%%%%%%%%%%%%%%%%%%%%%%%%%%%%%%%%
%%%%%%%%%%%%%%%%%%%%%%%%%%%%%%%%%%%%%%%%%%%%%

%%%%%%%%%%%%%%%%%%%%%%%%%%%%%%%%%%%%%%%%%%%%%
%%%%%%%%%%%%%%%%%%%%%%%%%%%%%%%%%%%%%%%%%%%%%
\section{Simulated samples}
\label{sec:sim}

The studies reported in the present document use simulated proton-proton collisions at $\sqrt{s}=14$~TeV and with pile-up of 200.
Specifically, 1000 $t\bar{t}$ events produced via gluon fusion or $q\overline{q}$~annihilation are generated using
PYTHIA8~\cite{Pythia1,Pythia2}. To model the pile-up, soft processes including inelastic soft QCD interactions
are generated using PYTHIA8 and the A3~tune~\cite{ATL-PHYS-PUB-2016-017}. The response of the {\it Generic Detector},
a simple tracking detector inspired by initial layouts for the ITk detector that ATLAS will use at the HL-LHC, is simulated
using the fast simulation inside the ACTS framework~\cite{Ai:2019kze}. The fast simulation implemented in the current version of ACTS
does include energy loss (via multiple scattering and bremsstrahlung). Secondaries are not propagated through the detector and nuclear
interactions are not simulated.

The {\it Generic Detector} has also been used
to simulate the dataset for the TrackML Challenge~\cite{Kiehn:2019tbl}. It is shown in Figure~\ref{fig:GenericDetector}.
The active detector volumes (bold solid lines) are arranged in successive layers that are parallel or perpendicular to the
beam line. The layers are grouped into nine volumes. Volumes 8, 13 and 17 in the centre of the detector are referred to as {\it barrel},
the remaining volumes form two {\it endcaps}. When a charged particle traverses an active volume, it leaves a space-point {\it hit}.
The layers are built from large numbers of silicon {\it modules}, and the total number of modules in the {\it Generic Detector} is 18728.

\begin{figure}[!htb]
    \centering
    \includegraphics[width=0.8\textwidth]{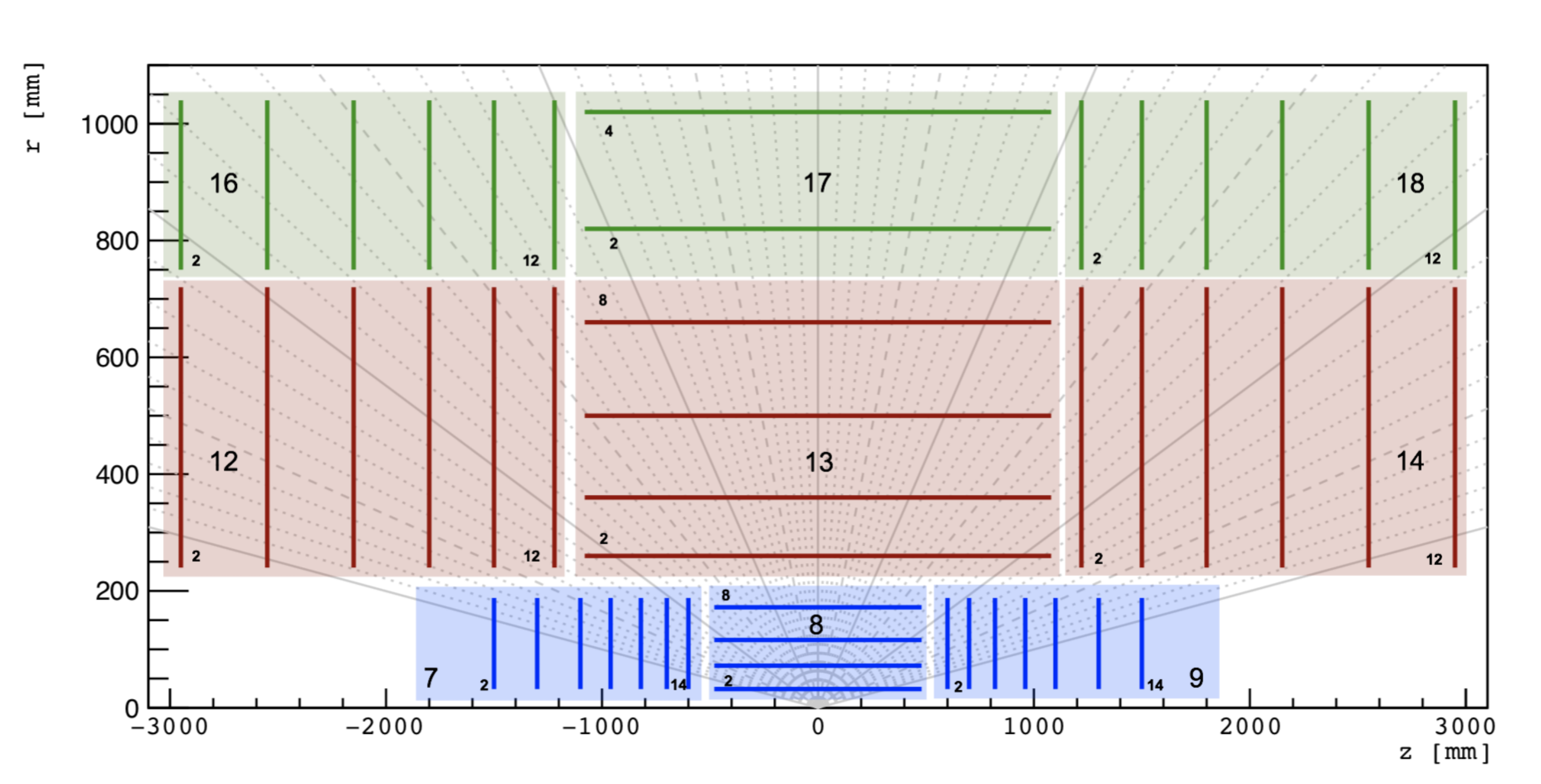}
    \caption{Layout of the {\it Generic Detector} defined in the ACTS framework. Figure taken from Ref.~\cite{Kiehn:2019tbl}.}
    \label{fig:GenericDetector}
\end{figure}

%\addcontentsline{toc}{section}{Simulation}
%%%%%%%%%%%%%%%%%%%%%%%%%%%%%%%%%%%%%%%%%%%%%
%%%%%%%%%%%%%%%%%%%%%%%%%%%%%%%%%%%%%%%%%%%%%

%%%%%%%%%%%%%%%%%%%%%%%%%%%%%%%%%%%%%%%%%%%%%
%%%%%%%%%%%%%%%%%%%%%%%%%%%%%%%%%%%%%%%%%%%%%
\section{Novel algorithm for graph creation}
\label{sec:data}

\subsection{Graph neural networks and graph representation of tracking data}

A GNN model acts on data represented by a graph. A graph is defined by a set of \textit{nodes} $\{V\}$ and by a set of \textit{edges} $\{E\}$. An edge is a connection between two nodes. The data from a tracking detector for a given event and their interpretation can be represented using a graph: a node represents a hit, and the existence of an edge between two nodes means that the nodes could potentially represent two successive hits on a track.
At the HL-LHC, $O(10^{5})$ hits per $t\bar{t}$ event are expected. A fully connected graph of such an event would have $O(10^{10})$ edges, most of them representing unphysical connections. A key feature of graph construction is therefore the choice of the edges.
A graph density~\cite{graphdensity} is defined as the ratio of the number of edges and the maximum possible number of edges. For directed graphs it is given by $\frac{n_E}{n_V \times(n_V - 1)} $, where $n_E$ and $n_V$ are the number of edges and nodes.\par

In Ref.~\cite{exatrkxpaper}, only the central region (barrel) of the tracking detector is considered. To avoid the problem with GPU memory consumption during the training of the GNN that will be discussed in Sec.~\ref{sec:memory}, the barrel is divided into $\mathrm{2\times8}$ sections of equal dimensions in $[\eta,\phi]$. A separate graph is constructed for each section and particle tracking is performed separately in each section. Connections between hits can only occur between successive detector layers and only one hit per layer per track is used\footnote{Two hits in a given layer occur in regions where neighbouring modules overlap. Such overlaps necessarily exist in a hermetic detector; they are useful for {\it in situ} measurements of the relative positions of the modules ({\it alignment}).}. Cuts based on the relative position of hit pairs are used to significantly reduce the number of edges.
Applied on the sample from Sec.~\ref{sec:sim}, this procedure leads to directed\footnote{In a directed graph, an edge has a direction; it can be thought of as an arrow pointing from one of the two hits to the other one. Here two edges (in opposite directions) are created for potential track connections.} graphs with $1900$ nodes and $12400$ edges on average, which corresponds to a density of $3.4\times10^{-3}$. This method is unsuitable for an extension to the endcaps because of their complex geometry. It suffers from the large combinatorics of pairs of hits in successive layers. A generic method also needs to take into account holes on tracks (hits of a track missing on a layer).
In Ref.~\cite{Choma:2020cry} a general graph construction approach is presented. Hit positions are projected into a new Euclidian space in which pairs belonging to the same particle are nearby, and pairs belonging to different particles are further apart. An unsupervised clustering algorithm is then applied to connect nodes.\par

\subsection{Learning any complex detector geometry}
\label{sec:learninggeom}

We propose a new, simple method for graph construction. It takes into account the complex geometry of realistic detectors as well as inefficiencies of the sensors\footnote{provided that they are included in the detector simulation}. The basic idea is to build graphs starting from a list of possible connections from one module to another. A connection is considered possible if tracks produced in a proton-proton collision can have successive hits on the two modules. Given that individual modules are small, this addresses the issue with combinatorics discussed above\footnote{An efficient implementation of this algorithm benefits from an event record in which the hits are already ordered by module. The algorithm can in principle be generalised to connections between {\it zones} other than modules.}. The list is established using simulated events. All hits of a given simulated particle are ordered in time and the connection between the modules of each pair of successive hits is included in the list. Entries in the list are directional (``the first module in the entry was hit before the second module''). Only simulated particles with $p_T~>~0.5~\GeV$ leaving at least 3 hits in the detector are considered. \par

This method, implemented in the ACTS framework, only needs to be run once to learn the geometry of a given detector. Using the full sample of $t\bar{t}$ events, 239699 module connections are found among the 18728 modules. \par

Without any further cuts on the edges, an event graph comprises $O(10^{7})$ edges. Technically speaking, we create directed graphs, using the direction from the list of module connections. Given that particles tend to move away from the interaction region, this effectively reduces the number of edges by a factor of two compared to the procedure discussed in the previous subsection. The direction of the edges will be ignored by the GNN presented in Sec.~\ref{sec:selections}, and taken into account by the track reconstruction presented in Sec.~\ref{sec:trackreco}. \par

\subsection{Further selection requirements}

To further reduce the number of edges, cuts are applied on the geometric parameters $z_0$, $\phi_{slope}$, $\Delta\phi$ and $\Delta\eta$. The parameter $z_0$ is defined as $\mathrm{z_{0}=z_{h_{1}}-r_{h_{1}}\times(\Delta z/\Delta r)}$ with $\mathrm{h_{1,2}}$ being the hits connected by the edge, $\Delta z$ the distance in~$z$ between the hits and $\Delta r$ their radial distance. The parameter $\phi_{\rm slope}$ describes the slope in $\phi$ of the edge with respect to the beam axis, and $\Delta\phi$, $\Delta\eta$ are respectively the difference in $\phi$ and $\eta$ of the hits connected by the edge. \par

The cuts are automatically adjusted, separately for each module connection, such that no true edge is rejected in the sample of 1000 $t\bar{t}$ events. Table~\ref{tab:cutflow} shows the mean number of edges per graph after each cut. Graphs created with this procedure have $1.0 \times 10^5$ nodes and $1.05 \times 10^6$ edges on average, which corresponds to a density of $1.0 \times 10^{-4}$. The mean number of true edges is $4.6\times10^{4}$.\par

\begin{table}[!hbt]
  \caption{Cutflow showing the mean number of edges per graph after the successive application of the cuts. The ``efficiency''~$\varepsilon$, defined as the ratio of the mean number of edges after the cuts over the mean number of edges before application of any cuts, is also shown.}
\begin{center}

\begin{tabular}{lll}
\multicolumn{1}{l}{}                                                    & Mean number of edges   & $\varepsilon$ \\ \hline
\multicolumn{1}{l|}{Initial}                                             & $18.5\times10^{6}$ & ----       \\
\multicolumn{1}{l|}{$z_{0}$}                                             & $3.56\times10^{6}$  & 19\%       \\
\multicolumn{1}{l|}{$z_{0}$, $\phi_{slope}$}                             & $1.32\times10^{6}$ & 7.1\%        \\
\multicolumn{1}{l|}{$z_{0}$, $\phi_{slope}$, $\Delta\phi$}               & $1.10\times10^{6}$  & 5.8\%      \\
\multicolumn{1}{l|}{$z_{0}$, $\phi_{slope}$, $\Delta\phi$, $\Delta\eta$} & $1.05\times10^{6}$ & 5.6\%      \\ \hline
\end{tabular}
\label{tab:cutflow}
\end{center}

\end{table}

\subsection{Attaching geometric and target information to the nodes and edges}

The hit coordinates in $(r, \phi, z)$ space are attached to each node as {\it features}. For each edge, the observables $(\Delta \eta, \Delta \phi, \Delta r, \Delta z)$ are computed from the hit coordinates of origin and destination node, and they are attached to the edge as features. As the GNN is trained using supervised learning, a target graph is associated to each input graph. The target graph is identical to the input graph, except that the nodes do not have any features and that a truth flag is the only feature attached to each edge.\par

%\addcontentsline{toc}{section}{Data}
%%%%%%%%%%%%%%%%%%%%%%%%%%%%%%%%%%%%%%%%%%%%%
%%%%%%%%%%%%%%%%%%%%%%%%%%%%%%%%%%%%%%%%%%%%%

%%%%%%%%%%%%%%%%%%%%%%%%%%%%%%%%%%%%%%%%%%%%%
%%%%%%%%%%%%%%%%%%%%%%%%%%%%%%%%%%%%%%%%%%%%%
\section{Training the graph neural network}
\label{sec:selections}

\subsection{Graph neural network model}

The GNN model we use closely follows the one from Ref.~\cite{exatrkxpaper}.
It has an \textit{encode-process-decode} architecture.
At the \textit{encode} stage, an \textit{encoder} transforms, separately for each node and each edge, the features attached to the node or edge into a 
\textit{latent} representation in a $D-$dimensional space. 
The \textit{encoder} is implemented using two fully-connected neural networks~(NNs): one for the node features and one for the edge features.
At the \textit{process} stage, an interaction network~\cite{InteractionNetwork,battaglia2018relational} is applied to update the latent features of the graph. This is achieved in two steps.
An \textit{edge network}, a fully-connected NN, updates the latent features of each edge taking into account the latent features of the edge and of the origin and destination nodes.
A \textit{node network}, a fully-connected NN, then updates the latent features of each node taking into account the latent features of the node and an aggregate of the latent features of its incoming \textit{and} outgoing edges. Including the latent features of the outgoing edges is needed given the construction of the graphs (cf. Sec.~\ref{sec:learninggeom}).
The interaction network is applied iteratively $L$~times, which propagates information through the graph. 
At the \textit{decode} stage, a \textit{decoder} implemented using a fully-connected NN
transforms the latent features of each edge into a classification score for the edge. 
This classification score is a measure of the probability that the edge represents a segment of a particle track.

\subsection{Training of the model and memory consumption during the training}
\label{sec:memory}

The training procedures used in ML are typically based on some form of gradient descent. When the number of model parameters is large, automatic differentiation in reverse mode~\cite{AutoDiff}
is used as the mainstay technique since it allows the computation of the full gradient (w.r.t. all model parameters) in one single pass~\cite{AutoDiff}. The advantages of this approach
come with the cost of large storage requirements that are proportional to the number of operations in the evaluated function~\cite{AutoDiff}. This is a limiting factor for the training of
GNNs that use a large~$L$ and that act on large graphs.

To avoid GPU memory exhaustion and allow training on large graphs, the GNN is trained using the TFLMS algorithm~\cite{TFLM} with TensorFlow~2.1.
With TFLMS, tensors can be temporarily swapped to the host when they are not immediately needed\footnote{An alternative or complementary approach would be to use gradient checkpointing~\cite{DBLP} to reduce the amount of intermediate results stored in memory for automatic differentiation, at increased cost in compute power.}.
Training is performed on a host equipped with two Nvidia \RTX\ GPUs\footnote{This GPU type is boosted for single precision calculations (32~bits), and we have checked that going from double precision to single precision does not affect the performance of the trained GNNs.} with 48 GB of memory each. The host is further equipped with two AMD EPYC 7262 processors with 8 cores each
and with 1024 GB of memory. \par

A simple GNN model like in Ref.~\cite{exatrkxpaper} ($L = 8$, $D = 128$, two layers in each of the NNs) is trained using 80~events for training and 20~events for validation.
A binary cross-entropy loss function is used.
Both classes in the loss function are assigned a weight. This weight is set to $1.0$ for true edges and to $0.1$ for fake edges.
The Adam optimiser~\cite{Adam} is used with a fixed learning rate of $5 \times 10^{-4}$.
The weights as well as the learning rate have been chosen based on a grid search.
The model is trained for 82400 iterations\footnote{One iteration uses one graph, i.e. one event.}, which take 5 days and 20 hours to be completed using a single \RTX. The full memory capacity of the GPU and 70~GB of resident memory on the host are used during the training. \par

The training of the GNN model can be accelerated using multiple GPUs and data parallelism.
This approach has been tested successfully using the Horovod package~\cite{sergeev2018horovod} and
two \RTX\ GPUs in a common host. As expected, the memory usage on the host approximately doubled as
both GPUs independently work on separate events, and both GPUs swap to the host.

%\addcontentsline{toc}{section}{GNN}
%%%%%%%%%%%%%%%%%%%%%%%%%%%%%%%%%%%%%%%%%%%%%
%%%%%%%%%%%%%%%%%%%%%%%%%%%%%%%%%%%%%%%%%%%%%

%%%%%%%%%%%%%%%%%%%%%%%%%%%%%%%%%%%%%%%%%%%%%
%%%%%%%%%%%%%%%%%%%%%%%%%%%%%%%%%%%%%%%%%%%%%
\section{Results}
\label{sec:results}

\subsection{Per-edge performance on the full detector}
\label{sec:peredgeperf}

Figure~\ref{fig:perfedges} shows the performance of the GNN trained in Sec.~\ref{sec:memory} for edge classification. The performance is assessed using 100~events that are different from those used for training and validation.
Figure~\ref{fig:classification} shows the distribution of the edge classification score, separately for true edges (on the track of a generated particle) and fake edges.
A per-edge efficiency is defined as the ratio of the number of true edges passing a given cut on the classification score over the number of total true edges. A per-edge purity is defined as the ratio of the number of true edges passing the cut over the total number of edges passing the cut.
Table~\ref{tab:MLEfficiencyPurity} summarises the values of these performance metrics for different cut values.
In order to obtain a high track reconstruction efficiency, we chose a cut value that leads to a per-edge efficiency of about 98\%, i.e. a cut at 0.8 on the edge score is used in the
following\footnote{No attempt to adjust the cut value as a function of $\eta$ is made.}.\par

Figure~\ref{fig:eff_vs_pt} shows the per-edge efficiency as a function of the transverse momentum $p_T$ of the generated particle that has caused the edge. Figures~\ref{fig:eff_eta} and~\ref{fig:pur_eta} show the per-edge efficiency and purity as a function of $\eta$ of the origin node.\par

\begin{figure}[!htbp]
\begin{center}
\subfloat[Edge classification score]{\includegraphics[height=6.1cm,width=0.48\textwidth]{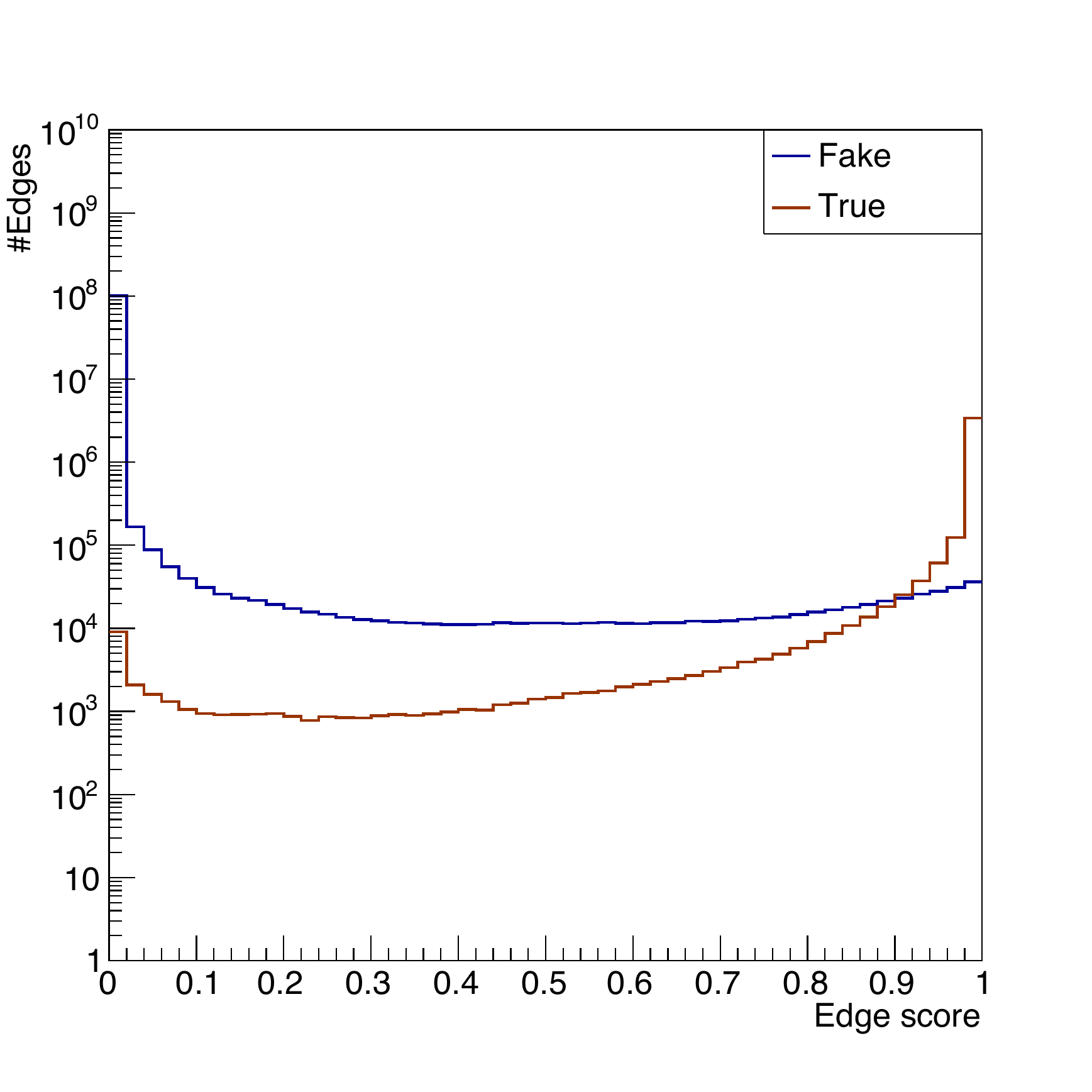} \label{fig:classification}}
\subfloat[Per-edge efficiency vs. $p_T$]{\includegraphics[width=0.48\textwidth]{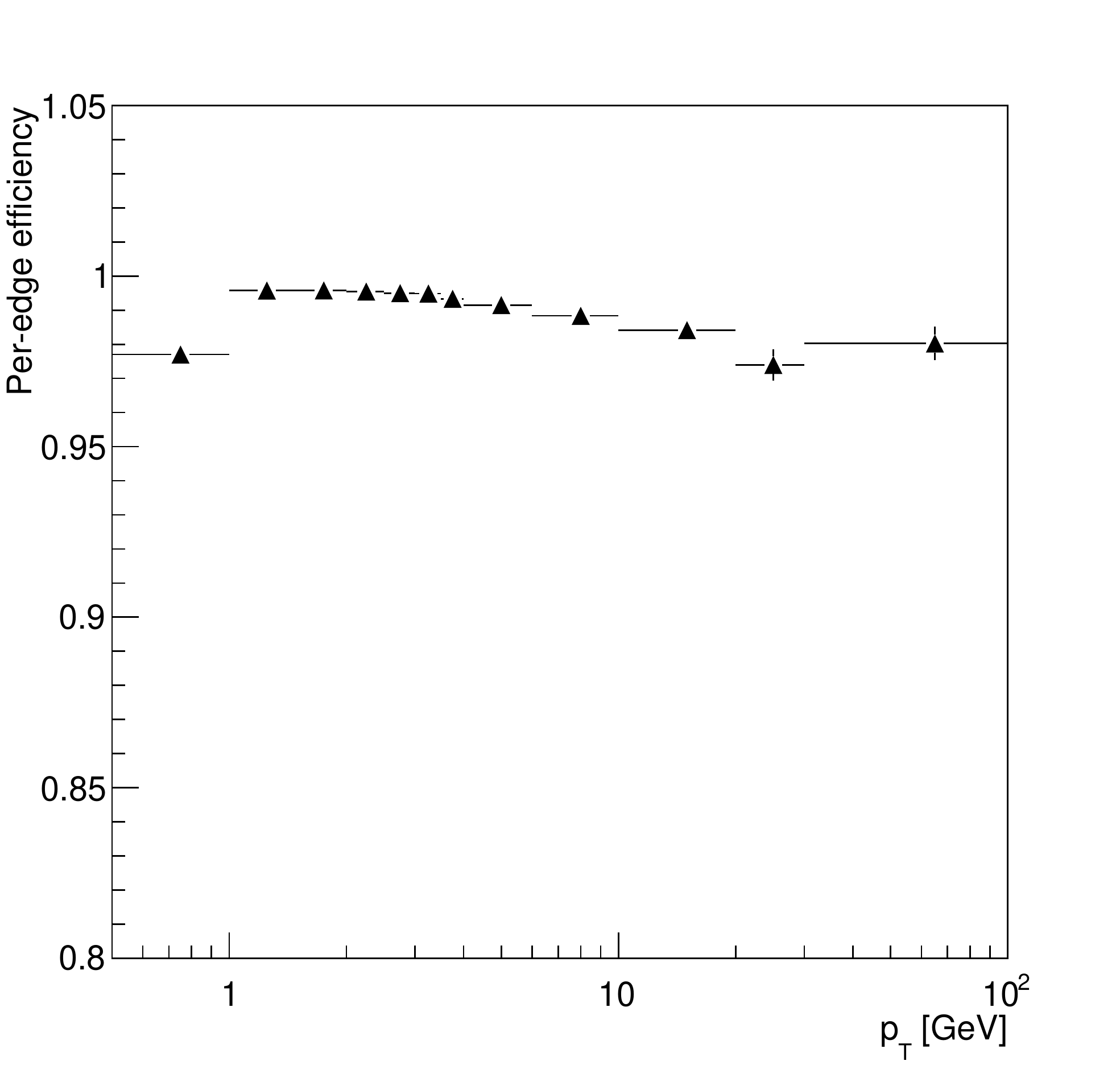} \label{fig:eff_vs_pt}} \\
\subfloat[Per-edge efficiency vs. $\eta$]{\includegraphics[width=0.48\textwidth]{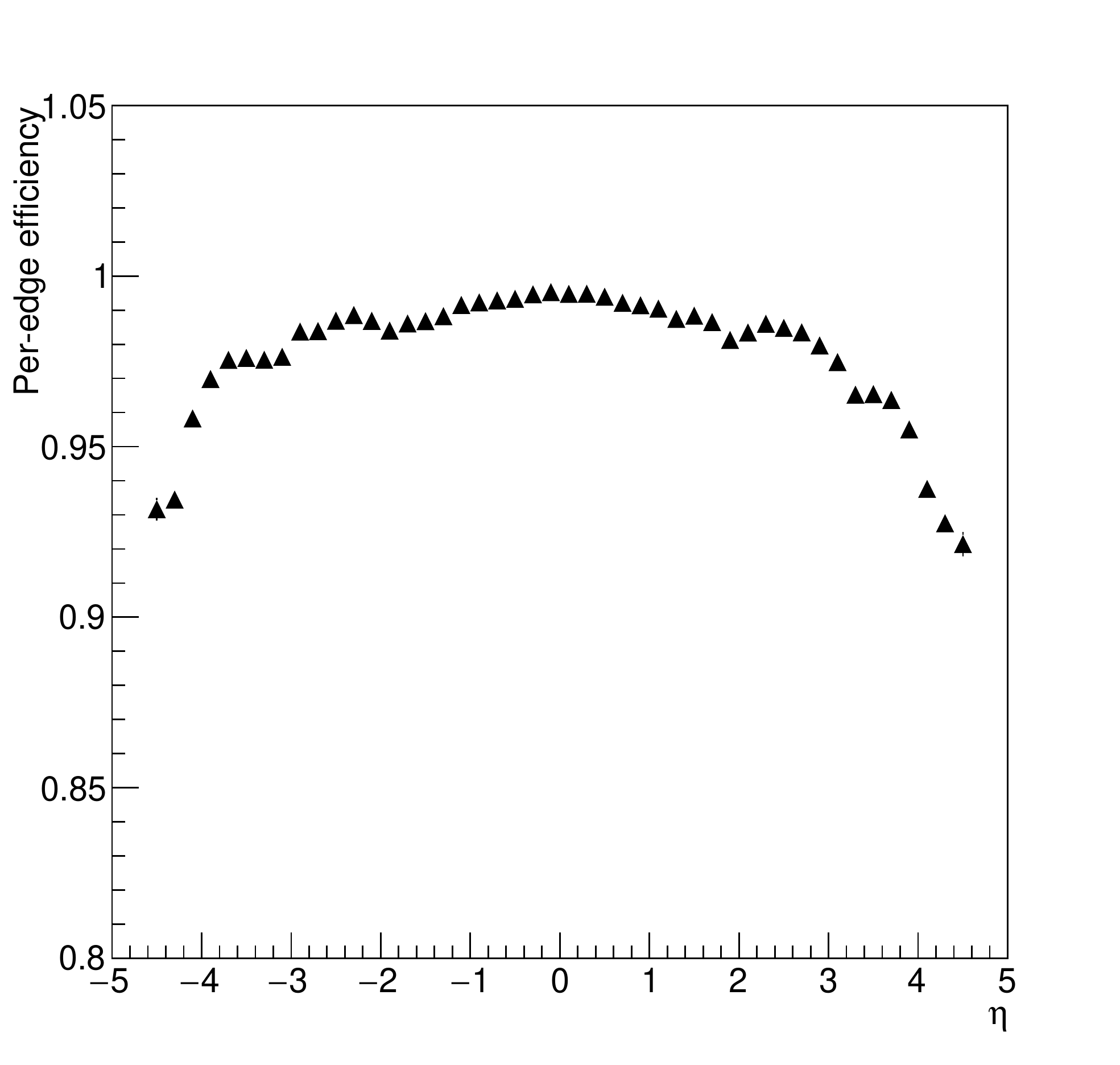} \label{fig:eff_eta}}
\subfloat[Per-edge purity vs. $\eta$]{\includegraphics[width=0.48\textwidth]{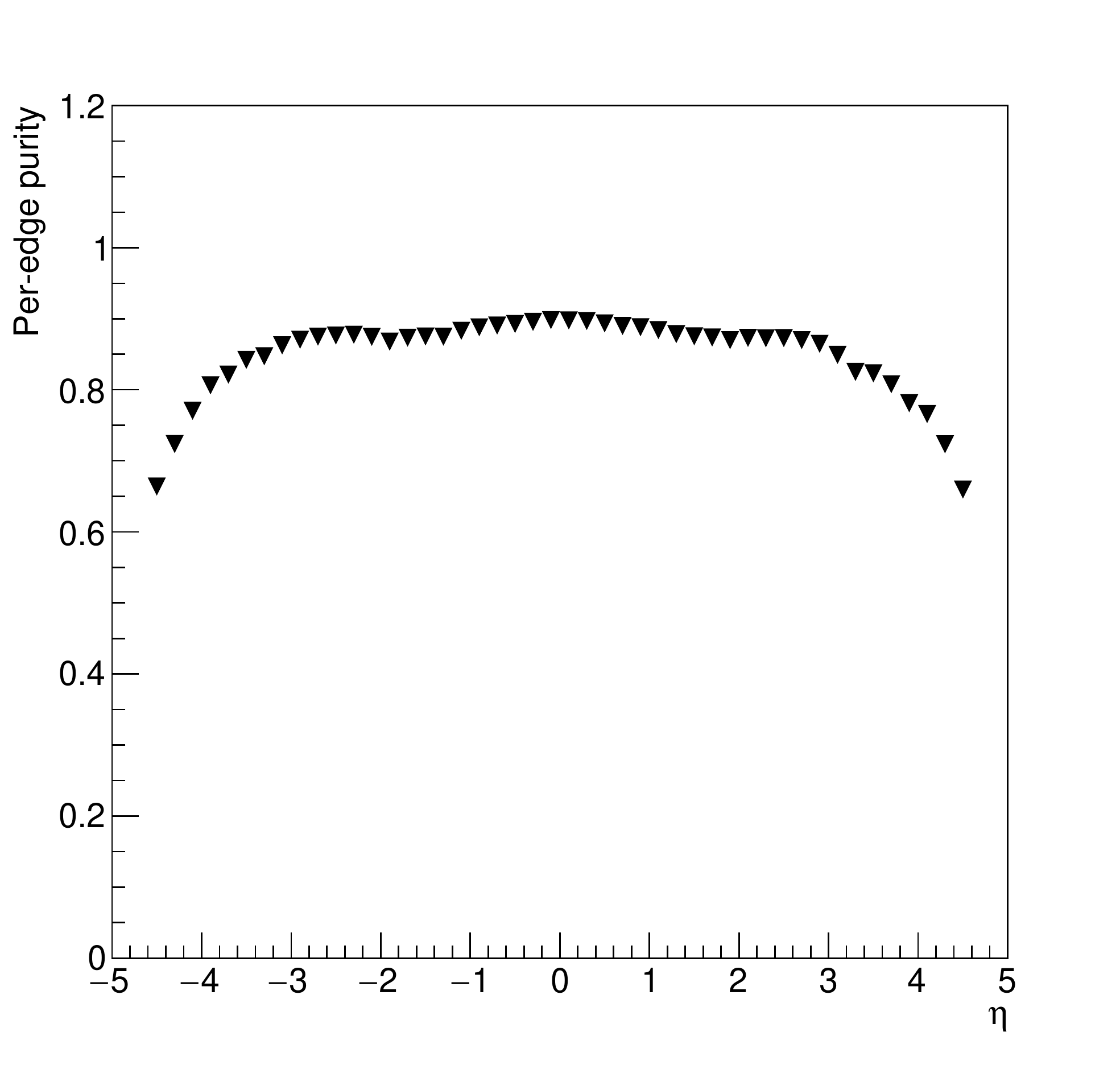} \label{fig:pur_eta}}\\
\caption{Performance of the GNN for edge classification on the full detector. The position in~$\eta$ (w.r.t. the centre of the detector) of the origin node is used to define $\eta$ for an edge.}
\label{fig:perfedges}
\end{center}

\end{figure}

\begin{table}[!hbt]
  \begin{center}
    \caption{Per-edge efficiency and purity for different cuts on the edge classification score.}
  \begin{tabular}{l|ccc}
    
\hline
Cut                          & 0.5       & 0.7       & 0.8       \\
\hline
Per-edge efficiency          & 0.992     & 0.987     & 0.982     \\
Per-edge purity              & 0.916     & 0.937     & 0.950     \\
\hline
\end{tabular}

\label{tab:MLEfficiencyPurity}
\end{center}
\end{table}

\subsection{Towards a more complex GNN model}

As discussed in Sec.~\ref{sec:memory}, a GNN training requires several days (or a lot of expensive hardware) to be completed. For studies of different GNN configurations with a fast turnaround time, we create smaller graphs (``{\it phi slice}'') by applying a cut $\phi\in]0,1[$. Graphs created with this cut comprise $O(10^{4})$ nodes and $O(10^{5})$ edges. Using this sample, the training of the model described above fits into the memory of the \RTX\ and also into the 32~GB of memory of an Nvidia \Tesla\ GPU\footnote{We mention this second GPU type because we have access to a relatively large number of these GPUs.}, and no swapping is needed. Overfitting is seen after 60000 iterations, which take 12 hours to be completed. We also test GNNs with deeper NNs using the phi slice graphs, e.g. one with 4 layers in each NN and $L = 12$, and one with 6 layers in each NN and $L = 12$. The training of the latter configuration requires the full memory of a \Tesla\ GPU plus 7 GB on the host for swapping. The training is complete after 90000 iterations, which take 1 day and 16 hours. \par

Figures~\ref{fig:slice_eff} and \ref{fig:slice_pur} show the per-edge efficiency and purity obtained on the phi slice graphs using different GNN models. More complex GNN models have the promise of even better performance than the GNN model discussed in Sec~\ref{sec:peredgeperf} above.

\begin{figure}[!htbp]
  
  \begin{center}
    
\subfloat[Per-edge efficiency vs. $\eta$]{\includegraphics[width=0.48\textwidth]{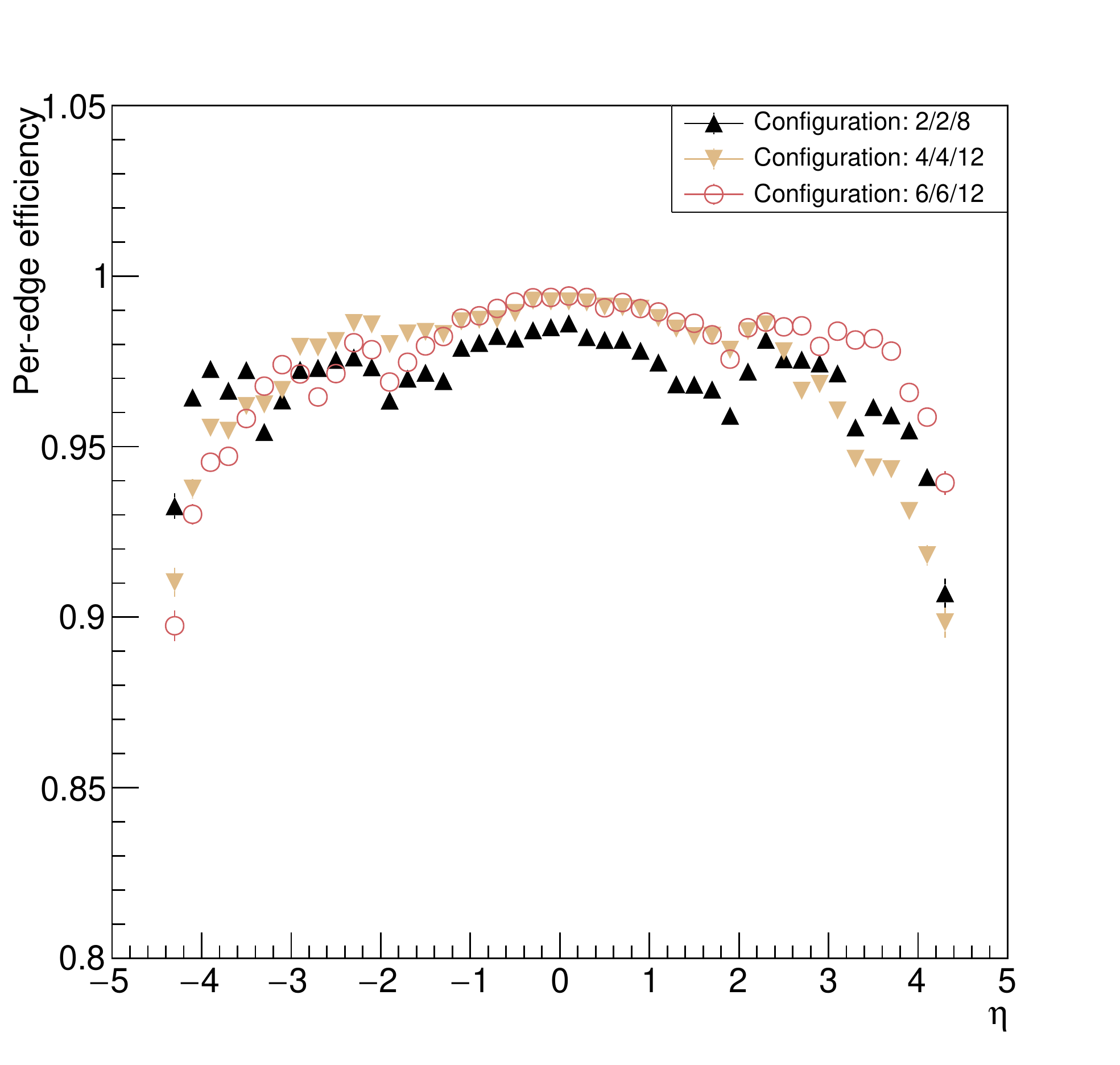} \label{fig:slice_eff}}
\subfloat[Per-edge purity vs. $\eta$]{\includegraphics[width=0.48\textwidth]{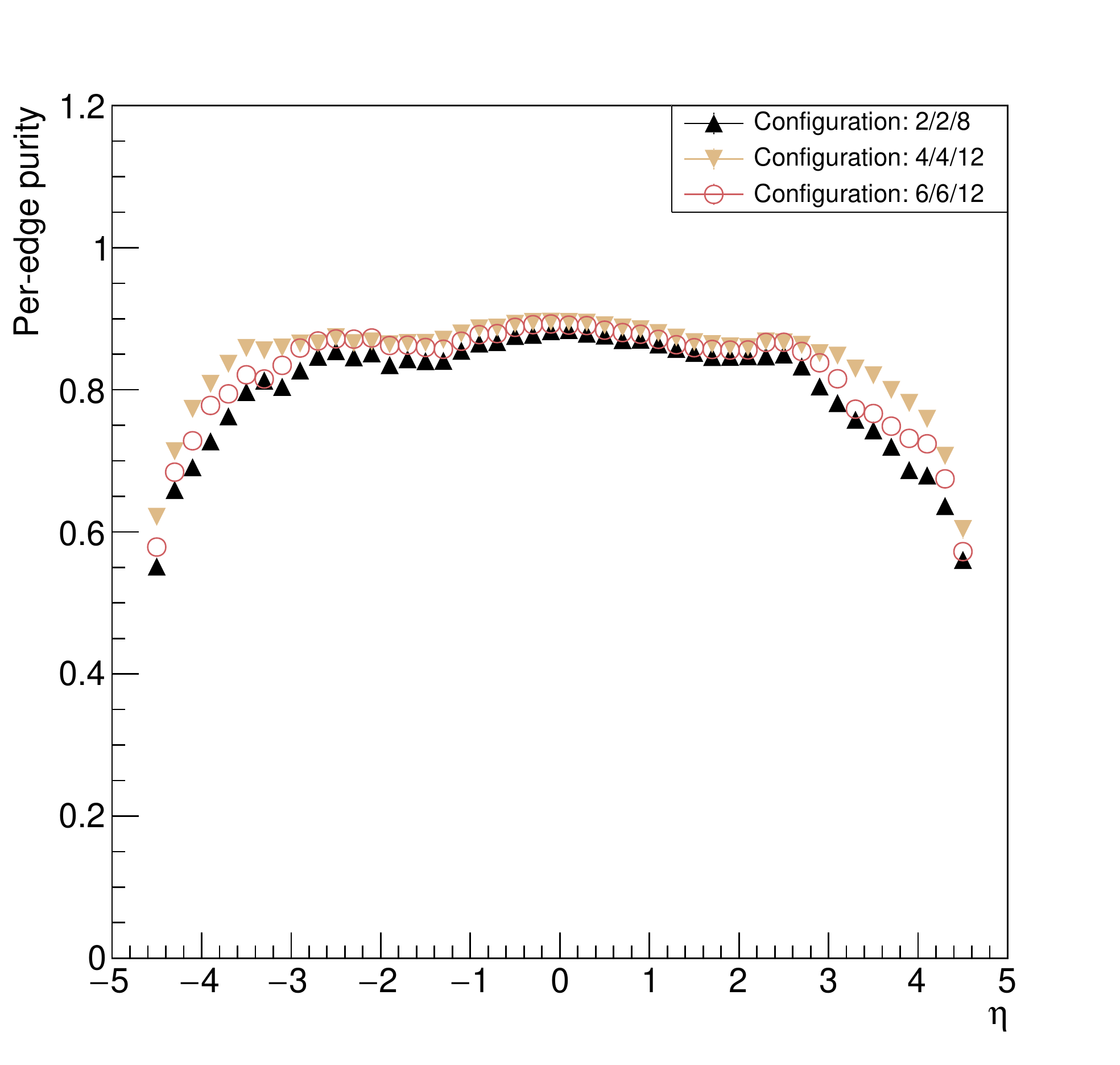} \label{fig:slice_pur}}\\
  \end{center}
  
  \caption{Performance of edge classification on the phi slice graphs. The notation ``2/2/8'' stands for a GNN with $L = 8$ and two layers in all of the NNs in the GNN.}
  
\end{figure}

\subsection{Track reconstruction efficiency on the full detector}
\label{sec:trackreco}

The next step is to build tracks starting from the graph and the edge scores. We have not performed any detailed studies of this step. We use a graph walk-through algorithm like the one in Ref.~\cite{exatrkxpaper}, and we augment it with a simple topological sort~\cite{manber1989iac}. In the walk-through only edges above a given threshold, set to 0.8, are kept. A topological sort of the nodes in the graph is performed before the walk-through. After this sort, origin hits tend to be listed before the destination hits. A hit can be associated with two tracks. In that case, the longest track is kept. This is to avoid cutting long tracks into multiple shorter ones. \par

To define a track reconstruction efficiency, a set of criteria to match generated particles to reconstructed tracks is needed. We define three sets. In the case of \textit{perfect} matching all hits from a given particle are on a track, and only these hits. For \textit{tight} matching, at least 90\% of the hits on a track come from the same particle, and at most 10\% of the hits from that particle are missing from the track. For \textit{loose} matching, at least 50\% of the hits on a track come from the same particle. \par

Figure~\ref{fig:reco_vs_eta} shows the efficiency to reconstruct generated particles with $p_T>1~\GeV$ that leave at least 3 hits in the detector. The efficiency is shown as a function of $\eta$ of the particle. The drop in the reconstruction efficiency with perfect and tight matching for $|\eta|\approx2$ corresponds to the  barrel-endcap transition region. Figure~\ref{fig:reco_vs_pt} shows the track reconstruction efficiency for generated particles that leave at least 3 hits in the detector, as a function of particle transverse momentum $p_T$. In general, barrel-endcap transition regions tend to be difficult, and some loss of performance compared to the central region is expected. The inefficiencies at large~$p_T$ are less expected. We note that they are less pronounced at the edge classification level, cf. Fig.~\ref{fig:eff_vs_pt}. \par

The results presented in Fig.~\ref{fig:trackeff} cannot be compared to those in Sec.~3.1.1 of Ref.~\cite{CERN-LHCC-2017-021}. The reference corresponds to a detector design report. The inclusion of inefficiencies caused by interactions in the detector material (including, but not limited to, nuclear interactions) in the results that are quoted in such a report is of the essence. Here we focus on algorithm development. We use a simplified simulation (cf. Sec.~\ref{sec:sim}) and we only consider particles that have left at least three hits in the detector.

\begin{figure}[!htbp]

  \begin{center}

    \subfloat[Track reconstruction efficiency vs. $p_T$]{\includegraphics[width=0.48\textwidth]{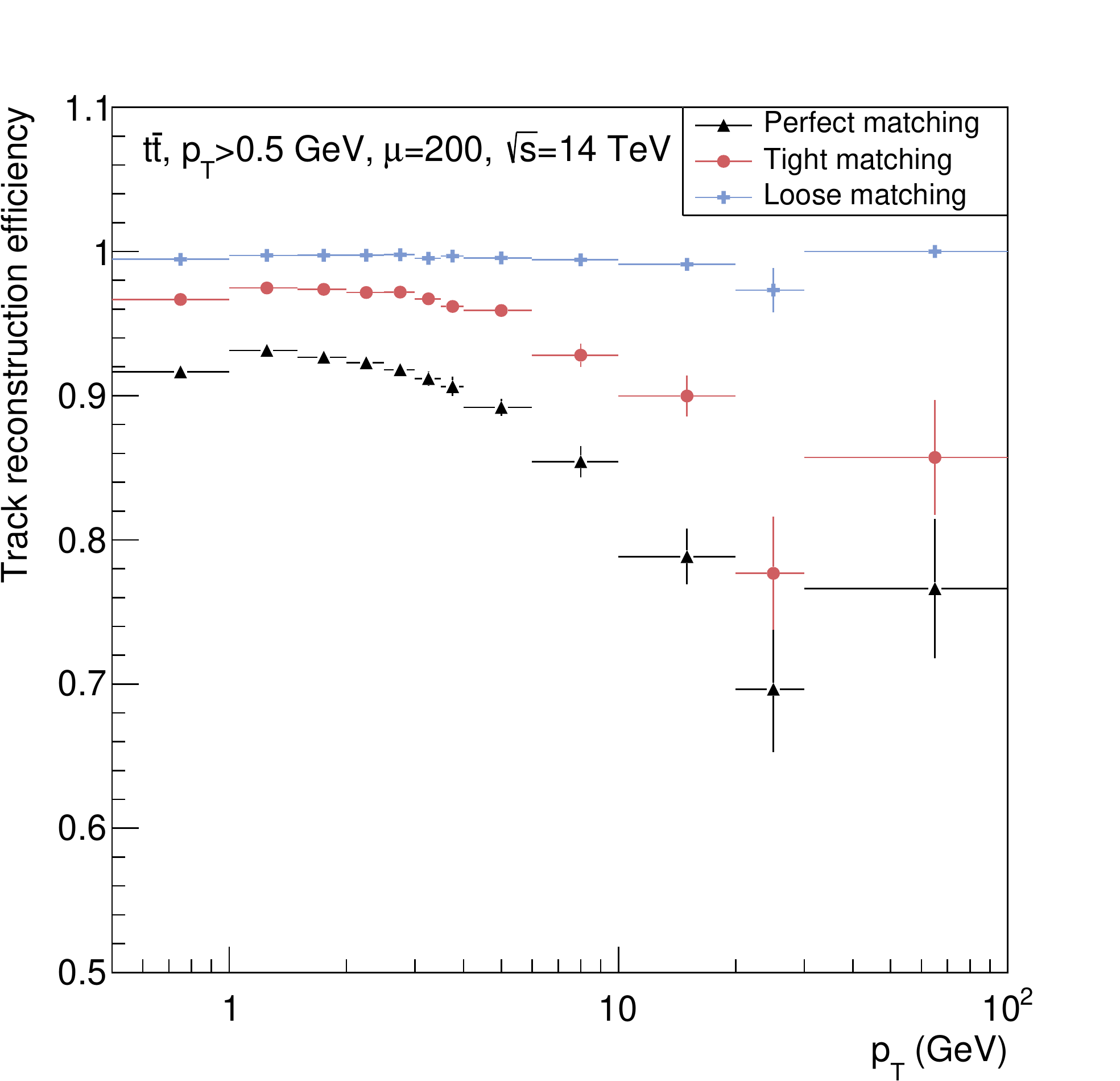} \label{fig:reco_vs_pt}}
    \subfloat[Track reconstruction efficiency vs. $\eta$]{\includegraphics[width=0.48\textwidth]{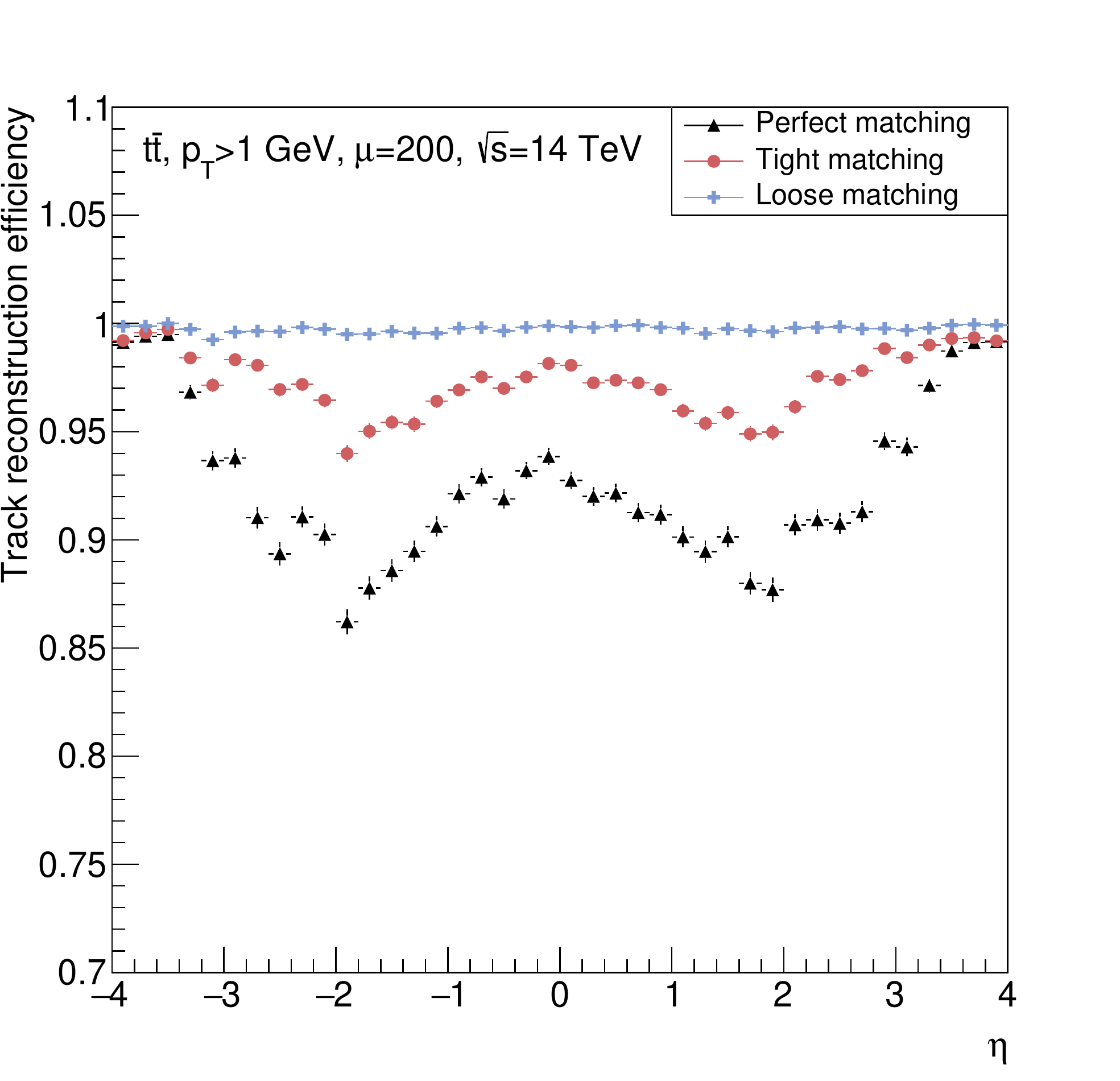} \label{fig:reco_vs_eta}}\\ 

  \end{center}

  \caption{Track reconstruction efficiency on the full detector, as a function of transverse momentum and pseudo-rapidity of the generated particle. Only particles that leave at least three hits in the detector are considered.}
  \label{fig:trackeff}
  
\end{figure}

%\addcontentsline{toc}{section}{Results}
%%%%%%%%%%%%%%%%%%%%%%%%%%%%%%%%%%%%%%%%%%%%%
%%%%%%%%%%%%%%%%%%%%%%%%%%%%%%%%%%%%%%%%%%%%%

%%%%%%%%%%%%%%%%%%%%%%%%%%%%%%%%%%%%%%%%%%%%%
%%%%%%%%%%%%%%%%%%%%%%%%%%%%%%%%%%%%%%%%%%%%%
\section{Conclusion and outlook}
\label{sec:concl}

We present results for track pattern recognition at the HL-LHC using GNNs. The architecture of the GNN itself closely follows 
that of earlier publications. The earlier work aims at establishing proof of principle and
uses algorithms that do not scale to the size nor to the complexity of a realistic detector. The developments presented here, including a novel
algorithm for graph construction and the use of advanced methods for memory management, make this scaling possible. Promising results for track
reconstruction in the full detector are obtained. The new algorithms are implemented in the ACTS framework. This makes it possible to
use them in simulation studies of more recent detector designs and to run them concurrently with other tracking algorithms on the same events.

%%%%%%%%%%%%%%%%%%%%%%%%%%%%%%%%%%%%%%%%%%%%%
%%%%%%%%%%%%%%%%%%%%%%%%%%%%%%%%%%%%%%%%%%%%%

\section*{Acknowledgements}
We thank our colleagues at the IN2P3 computing centre (CC-IN2P3) in Lyon~(Villeurbanne) for the smooth operation of
their GPU production platform, and for the successful deployment of a new experimental platform dedicated to machine learning
developments that require large amounts of memory.
Without these resources, the present studies would not have been possible.
We thank G\'erald Foliot (TGIR Huma-Num) for his contributions to the deployment of the new platform.

%%%%%%%%%%%%%%%%%%%%%%%%%%%%%%%%%%%%%%%%%%%%%
%%%%%%%%%%%%%%%%%%%%%%%%%%%%%%%%%%%%%%%%%%%%%
\begin{comment}
\appendix
\section{Some title}
Please always give a title also for appendices.
\end{comment}
%%%%%%%%%%%%%%%%%%%%%%%%%%%%%%%%%%%%%%%%%%%%%
%%%%%%%%%%%%%%%%%%%%%%%%%%%%%%%%%%%%%%%%%%%%%

%%%%%%%%%%%%%%%%%%%%%%%%%%%%%%%%%%%%%%%%%%%%%
%%%%%%%%%%%%%%%%%%%%%%%%%%%%%%%%%%%%%%%%%%%%%
%
% BibTeX or Biber users please use (the style is already called in the class, ensure that the "woc.bst" style is in your local directory)
% \bibliography{name or your bibliography database}
%
%%%\normalelm needed to make title in italic in references, break the gitlab comilation
\bibliography{refs_new} 

%\begin{thebibliography}{}
%Journal Author, Journal \textbf{Volume}, page numbers (year)
%Book Author, \textit{Book title} (Publisher, place, year) page numbers
% etc
%\end{thebibliography}

\end{document}

% end of file template.tex